\documentclass[letterpaper,twocolumn,showpacs,
prl 
,superscriptaddress
]{revtex4}

\usepackage{graphicx}
\usepackage{amssymb}
\usepackage{amsmath,amsfonts,latexsym}
\usepackage{array,tabularx,color}
\usepackage{dcolumn}               
\usepackage[normalem]{ulem}
\usepackage{comment}

\newcommand{\vect}[1]{{\mathbf #1}}
\newcommand{\Frac}[2]{\displaystyle\frac{#1}{#2}}

\begin{document}


\title{Spontaneous and triggered vortices in polariton OPO
  superfluids}

\author{F. M. Marchetti}
\affiliation{Departamento de F\'isica Te\'orica de la Materia
Condensada, Universidad Aut\'onoma de Madrid, Madrid 28049, Spain}

\author{M. H. Szyma\'nska}
\affiliation{Department of Physics, University of Warwick, Coventry,
  CV4 7AL, UK}
\altaffiliation{also at London Centre for Nanotechnology, UK}

\author{C. Tejedor}
\affiliation{Departamento de F\'isica Te\'orica de la Materia
  Condensada, Universidad Aut\'onoma de Madrid, Madrid 28049, Spain}

\author{D. M. Whittaker}
\affiliation{Department of Physics and Astronomy, University of
  Sheffield, Sheffield, S3 7RH, UK}

\date{March 25, 2010}       

\begin{abstract}
  We study non-equilibrium polariton superfluids in the optical
  parametric oscillator (OPO) regime using a two-component
  Gross-Pitaevskii equation with pumping and decay. We identify a
  regime above OPO threshold, where the system undergoes spontaneous
  symmetry breaking and is unstable towards vortex formation without
  any driving rotation. Stable vortex solutions differ from metastable
  ones; the latter can persist in OPO superfluids but can only be
  triggered externally. Both spontaneous and triggered vortices are
  characterised by a generalised healing length, specified by the OPO
  parameters only.
\end{abstract}

\pacs{42.65.Yj, 47.32.-y, 71.36.+c}



 
 







\maketitle

Since the first observation of stimulated
scattering~\cite{savvidis00:prl}, resonantly driven polariton
microcavities have been the subject of intensive research. Significant
advances have taken place towards a new generation of low threshold
lasers and ultrafast optical amplifiers and switches. However, only
very recently resonantly pumped polaritons have been shown to exhibit
a new form of non-equilibrium superfluid
behaviour~\cite{amo09,amo09_b,sanvitto09}. In the OPO
regime~\cite{stevenson00}, polaritons are continuously injected into
the \emph{pump} state and, above a pump strength threshold, undergo
coherent stimulated scattering into the \emph{signal} and \emph{idler}
states. Superfluidity has been tested through frictionless
flow~\cite{amo09} by triggering a traveling signal with an additional
pulsed probe laser. Moreover, metastability of quantum vortices and
persistence of currents have been proven by using a pulsed
Laguerre-Gauss beam~\cite{sanvitto09}. Vorticity has been observed to
be transferred into to the OPO signal and to persist in absence of the
driving rotating probe.

\begin{figure}
\begin{center}
\includegraphics[width=0.75\linewidth,angle=0]{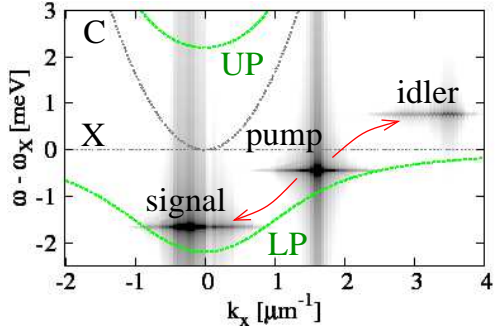}
\end{center}
\caption{(Color online) OPO spectrum for a top-hat pump of FWHM
  $\sigma_p=70\mu$m and intensity $f_p=1.24 f_p^{\text{(th)}}$, where
  $f_p^{\text{(th)}}$ is the threshold strength for OPO. Polaritons
  injected resonantly at $(k_p,0)$ and $\omega_p$ undergo coherent
  stimulated scattering into the signal and idler states, which are
  blue-shifted with respect to the bare lower polariton (LP)
  dispersion (green dotted line) because of interactions. Cavity
  photon (C) and exciton (X) dispersions are plotted as gray dotted
  lines.}
\label{fig:spect}
\end{figure}

The polaritonic system is intrinsically non-equilibrium: Continuous
pumping is needed to balance the fast polariton decay, of the order of
picoseconds, and maintain a steady state regime. In strong contrast
with equilibrium superfluids, which ground state is flowless, pump and
decay lead to currents that carry polaritons from gain to loss
dominated regions. As a result of the interplay between these currents
and a confining potential, polaritons non-resonantly injected into a
microcavity have been shown to become unstable to spontaneous
formation of vortices~\cite{lagoudakis08,nardin10} and vortex
lattices~\cite{keeling08}.
For resonant excitation, currents arise in the OPO regime due to the
simultaneous presence of pump, signal and idler emitting at different
momenta (see Fig.~\ref{fig:spect}).
In this Letter we identify a regime of the OPO, where remarkably, even
in the absence of disorder or trapping potentials, the system
undergoes spontaneous breaking of the system symmetry and becomes
unstable towards the formation of a quantized vortex state with charge
$m=\pm 1$. We show that these spontaneous stable vortex solutions are
robust to noise and to any other external perturbation, and thus
should be experimentally observable.
Spontaneous stable vortices differ from metastable ones, which can
only be injected externally into an otherwise stable symmetric state,
and which persistence is due to the OPO superfluid
properties~\cite{sanvitto09}. The metastable vortex is a possible but
not unique stable configuration of the system.
We find that shape and size of the metastable vortices are independent
on the external probe. In addition, like in equilibrium superfluids,
both stable and metastable vortices are characterised by a healing
length which is determined by the parameters of the OPO system
alone. Controlled creation of vortices in OPO has been recently
achieved by a weak continuous probe~\cite{krizhanovskii10}.
Metastable vortices have been also recently discussed for
non-resonantly pumped polariton condensates in Ref.~\cite{wouters09}.

\paragraph{Model}
We describe the OPO dynamics via Gross-Pitaevskii (GP)
equations~\cite{whittaker2005_b} for coupled cavity and exciton fields
$\psi_{C,X} (\vect{r},t)$ with pumping and decay($\hbar=1$):
\begin{multline}
  i\partial_t \begin{pmatrix} \psi_X \\ \psi_C \end{pmatrix}
  = \begin{pmatrix} 0 \\ F_p + F_{pb} \end{pmatrix}
\\ + 
  \begin{pmatrix} \omega_X -i \kappa_X + g_X|\psi_X|^2& \Omega_R/2 \\
  \Omega_R/2 & \omega_C -i \kappa_C \end{pmatrix} \begin{pmatrix}
    \psi_X \\ \psi_C
  \end{pmatrix}\; .
\label{eq:model}
\end{multline}
We neglect the exciton dispersion and assume a quadratic dispersion
for photons, $\omega_C=\omega_C^{0} -\frac{\nabla^2}{2m_C}$. The
fields decay with rates $\kappa_{X,C}$ and $\Omega_R$ is the Rabi
splitting.  The cavity field is driven by a continuous wave pump,
\begin{equation}
  F_p(\vect{r},t) = \mathcal{F}_{f_p,\sigma_p} (r) e^{i (\vect{k}_p
    \cdot \vect{r} - \omega_p t)}\; ,
\label{eq:pumpe}
\end{equation}
where $\mathcal{F}_{f_p,\sigma_p}$ is either a Gaussian or a top-hat
spatial profile with strength $f_p$ and full width at half maximum
(FWHM) $\sigma_p$. The exciton interaction strength $g_X$ can be set
to one by rescaling both fields $\psi_{X,C}$ and pump strength $F_{p}$
by $\sqrt{\Omega_R/(2 g_X)}$. For the simulations shown in this Letter
$m_C=2 \times 10^{-5} m_0$, the energy zero is fixed to
$\omega_X=\omega_C^{0}$ (case of zero detuning), $\Omega_R = 4.4$meV
and $\kappa_{X,C}$ are fixed so that to give a polariton lifetime of
$3$ps.

\begin{figure}
\begin{center}
\includegraphics[width=1.0\linewidth,angle=0]{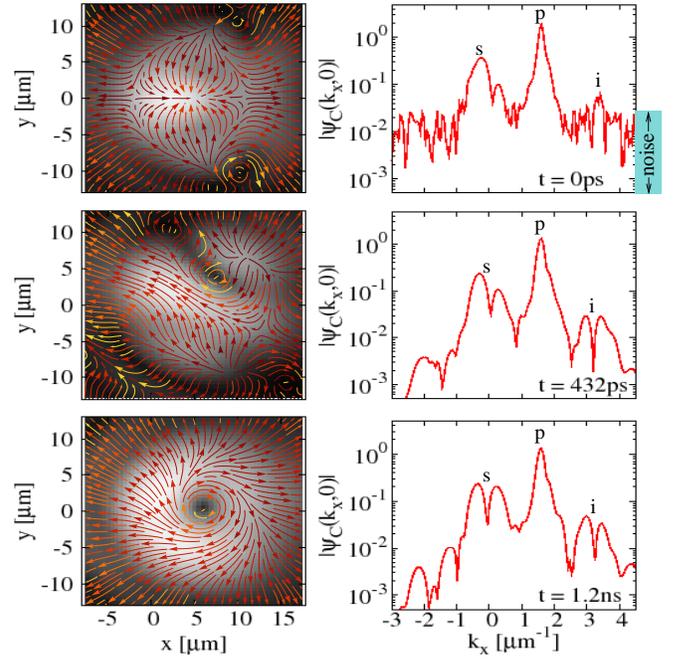}
\end{center}
\caption{(Color online) Filtered signal profile
  $|\psi_C^{s}(\vect{r},t)|$ with supercurrents $\nabla \phi_C^{s}
  (\vect{r},t)$ (left panels) and full momentum emission
  $|\psi_C(k_x,0,t)|$ (right in arb. units) at three different times:
  $t=0$ (first row), $t=432$ps (second) and $1.2$ns (third). The
  strength of the $\sigma_p=35\mu$m top-hat pump is $f_p=1.12
  f_p^{\text{(th)}}$. At $t=0$ a pulsed weak random noise of strength
  $0.01$ (see text) is added to the OPO steady state (first row) and
  at $t=432$ps a vortex, with $m=-1$, enters the signal and settles
  into a steady state. The noise strength only influences the time
  required for the vortex to appear.}
\label{fig:noise}
\end{figure}

%
For homogeneous pumps, the conditions under which a stable OPO
switches on can be found analytically by making use of the plane-wave
approximation~\cite{ciuti03,whittaker05}. However, for a finite size
pump~\eqref{eq:pumpe}, one has to resort to a numerical
analysis~\cite{whittaker2005_b}. Here, we numerically solve
Eq.~\eqref{eq:model} on a 2D grid by using both a
5$^{\text{th}}$-order adaptive-step Runge-Kutta algorithm and the
Crank-Nicholson method. Fixing the pump momentum $(k_p,0)$ close to
the LP inflection point, we find the pump strength threshold
$f_p^{\text{th}}$ for which signal and idler states get exponentially
populated. The OPO signal spatial profile $|\psi_{C,X}^{s}|
e^{i\phi_{C,X}^{s}}$ can be obtained by either filtering in a cone
around the signal momentum at a given time or by filtering the time
resolved spectrum in an interval around the signal energy (see
Fig.~\ref{fig:spect}). The photon component of the filtered OPO signal
profile $|\psi_C^{s}|$ and its supercurrent $\nabla \phi_C^{s}$ are
shown in the first left panel of Fig.~\ref{fig:noise}. We only select
OPO solutions which reach a steady state.
Note that the pump direction $(k_p,0)$ leaves the symmetry $y \mapsto
-y$ intact. This symmetry, while allowing vortex-antivortex pairs,
does not permit single vortices.

\begin{figure}
\begin{center}
\includegraphics[width=1.0\linewidth,angle=0]{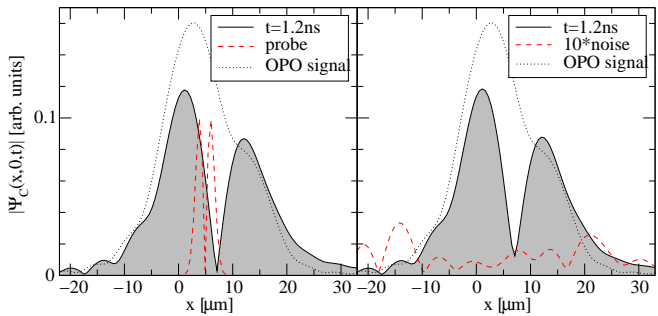}
\end{center}
\caption{(Color online) Steady state filtered signal profile (dotted
  line) $\psi_C^s (x,0,t)$ for $y\simeq 0$ before the arrival of
  either a vortex probe~\eqref{eq:probe} with $\sigma_{pb} \simeq
  1\mu$m (left panel, red dashed line) or a noise pulse of strength
  0.01 (right panel, red dashed line) --- same OPO conditions as
  Fig.~\ref{fig:noise}. After the arrival of any perturbation breaking
  the $y\mapsto -y$ symmetry, the same vortex with charge $m=\pm 1$
  (solid shaded curve) stabilises into the signal.}
\label{fig:profi}
\end{figure}

One can perform a dynamical stability analysis of the OPO by adding
small fluctuations to the steady state mean-field: The existence of
modes with positive imaginary part in the excitation spectrum
indicates dynamical instability towards the growth of different
modes. Equivalently, stability can be checked numerically by
introducing a weak noise. In particular, we add white noise as a quick
pulse to both modulus $|\psi_{X,C}(\vect{k},t)|$ and phase $\phi_{X,C}
(\vect{k},t)$ of excitonic and photonic wavefunctions in momentum
space. The noise added to the phase has amplitude $2\pi$, and for the
modulus we specify the noise strength in units of the maximum of the
pump intensity in momentum space.
%
\paragraph{Stable vortex solutions}
Remarkably, we have singled out steady symmetric OPO states, as shown
in the first row of Fig.~\ref{fig:noise}, which are unstable towards
the spontaneous formation of stable vortex solutions. Once the $y
\mapsto -y$ symmetry is broken by the noise pulse of any strength, we
have observed a vortex with quantised charge $m=\pm 1$ ($m=\mp 1$)
entering and stabilising into the OPO signal (idler). In the case of
Fig.~\ref{fig:noise} and the right panel of Fig.~\ref{fig:profi}, the
noise strength is $0.01$ and $432$ps after the noise pulse, a vortex
with $m=-1$ ($m=+1$) enters the signal (idler) and
stabilises. Different noise strengths do not affect the final steady
state, but only the \emph{transient} time the system needs to
accommodate the vortex and reach the steady configuration. Note that
parametric scattering constrains the phases of pump, signal and idler
by $2\phi_p=\phi_s+\phi_i$. Therefore a vortex in the signal implies
an antivortex in the idler and viceversa.

Further, we examine whether this vortex steady state is dynamically
stable by applying an additional noise pulse. For weak noise, with a
strength up to $0.1$, the vortex is stable and can only drift around a
little before settling again into the same state. For strong noise,
with strength $1$ and above, the vortex gets washed away, but after a
transient period, the very same state enters and stabilises again into
the signal, with the possibility of flipping vorticity, as discussed
later. Different noise strengths do not affect the final steady state,
but only the transient time.

Alternatively, one can break the $y \mapsto -y$ symmetry by a weak
pulsed vortex probe, and assess whether the stable steady state is in
any way dependent on the external perturbation.
As already shown in Refs.~\cite{whittaker07,sanvitto09}, vortices with
charge $m=\pm1$ can be generated in the OPO signal and idler, by
adding a Laguerre-Gauss pulsed probe:
\begin{multline}
  F_{pb}(\vect{r},t) = f_{pb} |\vect{r}-\vect{r}_{pb}|
  e^{-|\vect{r}-\vect{r}_{pb}|^2/(2\sigma^2_{pb})} e^{i m \varphi} \\
  \times e^{i (\vect{k}_{pb} \cdot \vect{r} - \omega_{pb} t)}
  e^{-(t-t_{pb})^2/(2\sigma^2_{t})} \; ,
\label{eq:probe}
\end{multline}
where the probe momentum $\vect{k}_{pb}$ and energy $\omega_{pb}$ are
resonant with either the OPO signal or idler state. The phase
$\varphi$ winds from $0$ to $2\pi$ around the vortex core
$\vect{r}_{pb}$. Shortly after the arrival of the pulsed probe
($\sigma_t=1$ps) there are two possible scenarios: either the vortex
is imprinted into the signal and idler and drifts around, or no vortex
gets transferred.
However, the homogeneous OPO states which are unstable towards the
spontaneous formation of stable vortices following a white noise
pulse, exhibit the same instability following a vortex probe pulse
(see left panel of Fig.~\ref{fig:profi}). The steady state vortex is
independent on both the probe intensity $f_{pb}$ and size
$\sigma_{pb}$, however the weaker the probe the longer the vortex
takes to stabilise, between $30$ and $400$ps for our system
parameters. The stable vortex following the Laguerre-Gauss probe is
exactly the same as the one triggered by a weak white noise (see right
panel of Fig.~\ref{fig:profi}), indicating that the probe acts only as
a symmetry breaking perturbation.

We can therefore infer that there are OPO conditions, as the one of
Fig.~\ref{fig:noise}, where the $y \mapsto -y$ symmetric solution is
dynamically unstable, and any symmetry breaking perturbation allows
the signal and idler to relax into a stable steady state carrying a
vortex with charge $\pm 1$. This suggests that such a vortex is the
genuine unique OPO stable state, which however cannot be accessed
without breaking the $y \mapsto -y$ symmetry. Instability of the
uniform state to spontaneous pattern (e.g., vortex) formation is a
typical feature of systems driven away from
equilibrium~\cite{cross93}. Similarly we find conditions for which the
uniform OPO solution is unstable to spontaneous formation of a
quantised vortex.

The system symmetry can also be broken by spatial disorder. Indeed, as
for non-resonantly pumped polaritons~\cite{lagoudakis08}, we have
found in our OPO simulations spontaneous stable vortices forming in a
disordered landscape.

%

%
\begin{figure}
\begin{center}
\includegraphics[width=1.0\linewidth,angle=0]{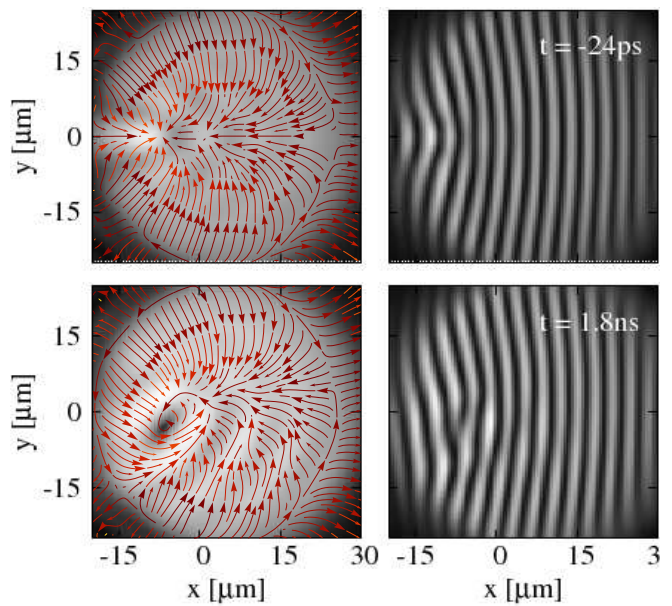}
\end{center}
\caption{(Color online) Generation of an $m=+1$ metastable vortex
  solution into the OPO signal ($f_p= 1.24 f_p^{\text{th}}$ and
  $\sigma_p=70\mu$m). First row: OPO filtered signal spatial profile
  together with currents (left) and interference fringes (right) are
  plotted at $t=-24$ps before the arrival of either a strong enough
  vortex probe~\eqref{eq:probe} or a strong enough noise pulse. The
  metastable vortex lasts for as long as our simulation (last row,
  $t=1.8$ns) and requires a threshold in the intensity of the
  perturbation breaking the $y \mapsto -y$ symmetry. The vortex
  appears in the interference fringes as a fork-like dislocation.}
\label{fig:topha}
\end{figure}
%
\paragraph{Metastability}
In addition to stable vortices, we have found OPO conditions
supporting metastable vortex solutions. In this case, the symmetric
OPO steady state is dynamically stable, but, because of its superfluid
properties, can support persistent metastable currents injected
externally. Metastable solutions can be equally induced by either a
vortex probe~\eqref{eq:probe} or a noise pulse. However, differently
from the stable case, such a solutions require a threshold in the
perturbation breaking the system symmetry. One example is shown in
Fig.~\ref{fig:topha}: the steady state shown in the second row can be
triggered by adding a white noise pulse to the symmetric OPO steady
state (first row of Fig.~\ref{fig:topha}) only for noise strengths
larger than 0.1. For weaker noises, the OPO signal is slightly
perturbed and rapidly goes back to its steady state vortexless
configuration. Alternatively, a vortex probe~\eqref{eq:probe} can also
be applied. For example, for a probe vortex with $\vect{r}_{pb} \simeq
(-6,-5)\mu$m, $\sigma_{pb}=4.5\mu$m, which is resonant with the OPO
signal momentum and frequency, a probe intensity $f_{pb} \ge
0.45f_{p}$ is required in order to generate the steady state vortex
shown in the last row of the figure. Right panels of
Fig.~\ref{fig:topha} show the interference fringes between signal and
pump state, obtained by considering the full emission in space
$|\psi_C(\vect{r},t)|$. A vortex corresponds to a fork-like
dislocation in the interferences.
 
The spatial position of both stable and metastable steady state
vortices is close to the position where the OPO signal has the
currents pointing inwards (see Figs.~\ref{fig:noise}
and~\ref{fig:topha}). We find stable vortices appearing quite close to
the pump threshold. At higher powers, the OPO tends to switch off in
the middle of the excitation spot and the system tends not to reach a
steady state. Finally, we checked that $m=\pm 1$ ($m=\mp 1$) vortex
solutions can appear only into the OPO signal (idler). A vortex probe
pulse of any charge $m$ injected resonantly to the pump momentum and
energy gets immediately transferred to an $m=\pm 1$ ($m=\mp 1$) vortex
in the signal (idler), leaving the pump vortexless.
The stability of signal $m>1$ vortices is outside the scope of this
work~\cite{szymanska10}.

\paragraph{Conservation of charge}
When generated by a noise pulse, stable and metastable vortices have
equal probability to be triggered with either charge
$\pm1$. Similarly, when vortices are triggered via a vortex probe,
their vorticity can flip during the transient period. Often, the
flipping is caused by the spontaneous appearance of two antivortices
at the edge of the signal, one recombining with the triggered vortex.
Topological charge inversion has been already predicted to occur in
confined atomic Bose-Einstein condensates at intermediate interactions
strengths~\cite{garciaripoll01}: In the presence of an asymmetry,
breaking rotation invariance, vorticity is not a conserved quantity.

\paragraph{Healing length} 
An approximate analytical expression for the vortex healing length can
be derived for homogeneous pumping, assuming that only signal and
idler carry opposite angular momentum $m$,
\begin{equation*}
  \psi^{s,i} (\vect{r}) = \sqrt{n_{s,i}} e^{i \vect{k}_{s,i} \cdot
    \vect{r}} e^{\pm i m \varphi} \Psi^{s,i} (r)\; ,
\end{equation*}
while the pump remains in a plane-wave state, $\psi^{p} (\vect{r}) =
\sqrt{n_{p}} e^{i \vect{k}_{p} \cdot \vect{r}}$, as supported by our
numerical analysis. For pump powers close to OPO threshold, it can be
shown that signal and idler steady state spatial profiles are locked
together and satisfy the following complex GP equation
\begin{equation*}
  \left[-\Frac{1}{2m_C} \left(\frac{d^2}{dr^2} + \frac{1}{r}
    \frac{d}{dr} - \frac{m^2}{r^2}\right) + \alpha \left(|\Psi^{s}|^2
    -1 \right)\right]\Psi^{s} = 0 \; ,
\end{equation*}
where $|\alpha| \simeq g_X \sqrt{n_s n_i}$. Keeping aside that
$\alpha$ is a complex parameter, this equation describes a vortex
profile~\cite{pitaevskii03} with healing length
\begin{equation}
  \xi = \left(2 m_C g_X \sqrt{n_s n_i}\right)^{-1/2}\; .
\label{eq:estim}
\end{equation}
This expression is similar to the one of an equilibrium superfluid,
with the condensate density replaced by the square root of the product
of the signal and idler densities. Further above threshold, the signal
and idler profiles are no longer locked together, and they start to
develop different radii. In both the simulations of
Figs.~\ref{fig:profi} and~\ref{fig:topha}, we find $\xi \simeq 4\mu$m,
compatible with the estimate~\eqref{eq:estim}. Recently, the
controlled creation of OPO vortices by a weak continuous
probe~\cite{krizhanovskii10} has allowed to experimentally test the
validity of Eq.~\eqref{eq:estim}.
%

%
To conclude, we have shown that close to the OPO threshold, the
polariton superfluid can spontaneously break the $y \mapsto -y$ system
symmetry and, even in absence of driving rotation, trapping, or
disorder potential, can be unstable towards the creation of a
quantised vortex state of charge $\pm 1$.
While the OPO symmetric state is generally stable, metastable vortices
can be injected by a strong enough external probe. The size of both
types of vortices is given by a healing length, which, close to
threshold, is analogous to the one of equilibrium superfluids.
Long lived metastable vortices have been experimentally realised in
Ref.~\cite{sanvitto09}, providing an evidence for existence of
persistent currents in this system. Since the parameters of our
simulations are close to those of current semiconductor microcavities,
the existence of stable spontaneous vortices in OPO should also be
within experimental reach. Moreover, these stable vortex states are
very robust to noise. Thus, apart from being yet another example of
exotic behaviour in polariton superfluids, they carry potential for
applications in, e.g., quantum information.
%

\acknowledgments We are grateful to J.J. Garc\'ia-Ripoll, D. Sanvitto,
J. Keeling, N.G. Berloff, and L. Vi\~na for stimulating
discussions. F.M.M. acknowledges financial support from the programs
Ram\'on y Cajal and INTELBIOMAT (ESF). This work is in part supported
by the Spanish MEC (MAT2008-01555, QOIT-CSD2006-00019) and CAM
(S-2009/ESP-1503).  We thank TCM group (Cavendish Laboratory,
Cambridge) for the use of computer resources.

%

\newcommand\textdot{\.}

\end{document}